\documentclass[12pt]{article}
\usepackage{amsmath,amssymb,bm}
\usepackage{graphicx}
\usepackage{cite}
\usepackage{geometry}
\newcommand{\be}{\begin{equation}}
\newcommand{\ee}{\end{equation}}
\newcommand{\bea}{\begin{eqnarray}}
\newcommand{\eea}{\end{eqnarray}}
\newcommand{\bref}[1]{(\ref{#1})}

\title{Infrared Divergence in QED and Fluctuation of Electromagnetic Fields}
\author{Takeshi Fukuyama\\
Research Center for Nuclear Physics (RCNP), Osaka University,\\
Ibaraki, Osaka 567-0047, Japan}
\date{}

\begin{document}
\maketitle

\begin{abstract}
We establish a no-go result for the infrared sector of quantum
electrodynamics.
Using the standard Fock-space formulation, we show that gauge invariance
enforces coherent soft-photon phases that guarantee the
Bloch--Nordsieck/Kinoshita--Lee--Nauenberg cancellation for all inclusive
observables.
The infrared divergences of perturbative amplitudes therefore do not signal
any physical instability of the theory, but reflect the universal quantum
dressing cloud inseparably accompanying charged particles.

We further demonstrate that the stochastic interpretation suggested by the
Schwinger--Keldysh effective action does not apply to four-dimensional Maxwell
theory.
Although infrared-sensitive imaginary terms appear in the SK effective action
and can be rewritten via a Hubbard--Stratonovich transformation, we prove that
conformal invariance forbids any infrared growth of these terms.
As a consequence, the associated auxiliary field cannot be interpreted as a
Langevin force, even in de~Sitter spacetime.

These results exclude infrared-induced classical stochastic dynamics for gauge
fields and clarify the physical distinction between QED and nearly massless
scalar fields in de~Sitter space.
\end{abstract}

\section{Introduction}

Infrared divergences (IRD) have long been a central conceptual feature of
quantum electrodynamics (QED) \cite{Berestetskii1, WeinbergQTF}.
In perturbative amplitudes involving charged particles, infinitely many
soft-photon modes contribute phase factors whose individual contributions are
logarithmically divergent.
The physical resolution of these divergences is well established:
in inclusive observables the singularities from virtual and real emissions
cancel through the Bloch--Nordsieck (BN) mechanism and its generalization by
Kinoshita, Lee, and Nauenberg (KLN) \cite{BN1937,Kinoshita,LN}.
This cancellation reflects a deep structural property of QED—namely, that
soft-photon phases form a correlated, universal cloud accompanying any charged
particle, with gauge invariance enforcing their coherence through the
Ward--Takahashi identity \cite{Ward1950, Takahashi1957}.

A complementary realization of this structure appears in the ``dressed''
asymptotic-state program initiated by Chung and Kibble and formulated
systematically by Faddeev and Kulish \cite{Chung, Kibble, FK1970}.
In this framework the asymptotic charged states are accompanied by coherent
soft-photon dressings, rendering the $S$-matrix explicitly infrared finite.
Although technically distinct from the traditional inclusive BN/KLN approach,
the Faddeev--Kulish construction embodies the same physical principle:
the soft sector does not constitute an independent dynamical degree of freedom,
but is inseparably tied to the charged state itself.

Recently, Morikawa \cite{Morikawa2016} analyzed infrared divergences using the
Schwinger--Keldysh (SK) in-in formalism
\cite{Schwinger1961,Keldysh1964,CalzettaHu1988}.
In that approach the imaginary part of the SK effective action encodes the
entanglement between long- and short-wavelength modes.
Through a Hubbard--Stratonovich (HS) transformation
\cite{Hubbard1959, Stratonovich1957},
this contribution may be rewritten in terms of an auxiliary classical field.
For nearly massless scalar fields in de~Sitter spacetime, the infrared sector
grows dynamically and long-wavelength modes effectively classicalize, allowing
this auxiliary field to be interpreted as a stochastic force.
Crucially, however, this mechanism relies on physical conditions that are absent
in four-dimensional Maxwell theory: conformal invariance forbids any infrared
enhancement of gauge-field fluctuations, and soft photons remain quantum
coherent.

In this work we demonstrate that these structural properties of QED exclude a
stochastic interpretation of infrared effects for gauge fields.
Using a Fock-space formulation, we show explicitly how gauge invariance enforces
coherent soft-photon phases that guarantee Bloch--Nordsieck and
Kinoshita--Lee--Nauenberg cancellations for all inclusive observables.
We further prove that, in four-dimensional Maxwell theory, the infrared-sensitive
imaginary terms appearing in the Schwinger--Keldysh effective action do not grow
in the infrared and therefore cannot generate physical noise.
As a consequence, the auxiliary field introduced by the HS representation
cannot be interpreted as a Langevin force, even in de~Sitter spacetime.
These results establish a no-go theorem: infrared divergences in QED do not give
rise to classical stochastic dynamics of gauge fields.

\section{Fock--space formulation and infrared fluctuation}

In the Fock-space formulation, the photon field is quantized with creation
and annihilation operators acting on the vacuum $|0\rangle$.  
The IR divergence appears in amplitudes
\[
\int_{k<\Lambda}\!\frac{d^3k}{2\omega_k}\,\bigl|{\cal M}(k)\bigr|^2 
\propto \ln(\Lambda/\lambda),
\]
but when inclusive probabilities are taken, interference among amplitudes
ensures the BN/KLN cancellation.  
Electromagnetic fluctuations arising from this coherence correspond to
quantum fluctuations of the Callen--Welton type \cite{Callen}, not to stochastic noise.
Gauge invariance, expressed by the Ward--Takahashi identity
$k_\mu \Gamma^\mu = 0$, enforces the correlated IR structure in both
virtual and real sectors \cite{Ward1950, Takahashi1957}.
It may be well established on the infrared divergence (IRD) and the
fluctuation of electromagnetic field in QED.
However, there are several ambiguities on the mutual relations.
In this letter we make clear the relation by studying the typical
diagrams comprehensively based on the conventional Fock space formalism.
It may also be useful to discriminate it from another formalism based on
the dressed state.

\begin{itemize}
\item {\bf Electron Propagator}

Firstly let us consider the radiative correction of mass operator which
is given by Fig.~1.
\begin{figure}[h]
\begin{center}
\includegraphics[scale=3.0]{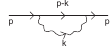}
\caption{Self energy diagram of electron.}
\label{fig:SE}
\end{center}
\end{figure}

The amplitude of this diagram is 
\be
-i\overline{M}(p)=(-ie)^2\int\gamma^\mu G(p-k)\gamma^\nu D_{\mu\nu}(k)
\frac{d^4k}{(2\pi)^4}.
\ee
After the regularization processes due to the double subtraction (see,
for instance, Section 119 of \cite{Berestetskii1}) and denoting it as
$M(p)$, we obtain 
\bea
M(p)&=&\frac{\alpha}{2\pi}(\gamma p-m)^2\Biggl\{
\frac{1}{2(1-\rho)}\left(1-\frac{2-3\rho}{1-\rho}\ln \rho\right)
\nonumber\\
&&\quad
-\frac{\gamma p+m}{m\rho}\left[
\frac{1}{2(1-\rho)}\left(2-\rho+\frac{\rho^2+4\rho-4}{1-\rho}\ln \rho\right)
+1+2\ln\frac{\lambda}{m}\right]\Biggr\},
\eea
with
\be
\rho=\frac{(m^2-p^2)}{m^2},\qquad
\gamma p=\sum \gamma_\mu p^\mu.
\ee
$m$ is an electron mass and $\lambda$ is a regulated photon mass.
The ultraviolet divergence and the infrared-sensitive terms appear in the
field renormalization
\begin{equation}
\psi_{\rm ren} = \sqrt{Z_1}\,\psi 
\end{equation}
with
\begin{equation}
Z_1 = 1 - \frac{\alpha}{2\pi}
\left(
  \frac{1}{2}\ln\frac{\Lambda^2}{m^2}
 + \ln\frac{\lambda^2}{m^2}
 + \frac{9}{4}
\right),
\end{equation}
where $\Lambda$ is the ultraviolet (UV) cutoff and $\lambda$ is an infrared
regulator (photon mass).
The UV divergence proportional to $\ln(\Lambda^2/m^2)$ is absorbed by
renormalization, whereas the infrared-sensitive term $\ln(\lambda^2/m^2)$
cannot be removed by wave–function renormalization alone.
In QED these infrared divergences cancel only after real soft-photon emission
is included in the inclusive probability, as ensured by the
Bloch–Nordsieck/KLN mechanism.
Thus the apparent IR divergence in the self-energy does not indicate a
physical inconsistency but reflects the universal soft-photon structure of
QED amplitudes.

\item {\bf Electron--Photon Vertex}

Secondly we consider the radiative correction of electron form factors
$\Gamma^\mu$,
\be
\Gamma^\mu=f(k^2)\gamma^\mu-g(k^2)\frac{\sigma^{\mu\nu}k_\nu}{2m},
\qquad
\sigma^{\mu\nu}=\frac{1}{2}(\gamma^\mu\gamma^\nu-\gamma^\nu\gamma^\mu),
\ee
where the two electron lines are external and the photon line is internal
(Fig.~2). In this case we can consider two channels: 
\be 
t=k^2 =(p_-+p_+)^2\leq 0~~\mbox{scattering channel}
\ee
and
\be
t \ge 4m^2 ~~\mbox{annihilation channel}.
\ee
The range $0\ge t\ge 4m^2$ is non-physical.
The radiative correction to the form factor $\Gamma^\mu$  is given by
Fig.~2.
\begin{figure}[h]
\begin{center}
\includegraphics[scale=2.0]{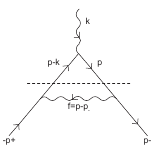}
\caption{The first vertex correction where the two electron lines are
external and the photon line is internal. That is, $k^2\neq 0$ but
$p_-^2=p_+^2=m^2$.}
\label{fig:vertex}
\end{center}
\end{figure}

Here 
\be
f(t)-1=\frac{\alpha t}{3\pi m^2}\left(\ln \frac{m}{\lambda}
-\frac{3}{8}\right),
\ee
\be
g(t)=\frac{\alpha}{2\pi},
\label{magnetic}
\ee
for $|t|=|k^2| \ll 4m^2$ \cite{Schwinger}, and
\be
f(t)-1=-\frac{\alpha}{2\pi}\left(\frac{1}{2}\ln^2\frac{|t|}{m^2}
+2\ln\frac{m}{\lambda}\ln \frac{|t|}{m^2}\right)+\left\{
\begin{array}{cc}
 i\frac{\alpha}{2}\ln\frac{t}{\lambda^2} & (t\gg 4m^2) \\
 0 & (-t\gg 4m^2)
\end{array}
\right.
\label{f}
\ee
\be
g(t)=-\frac{\alpha m^2}{\pi t}\ln\frac{|t|}{m^2}+\left\{
\begin{array}{cc}
 i\frac{\alpha m^2}{t} & (t\gg 4m^2) \\
 0 & (-t\gg 4m^2)
\end{array}
\right.
\label{g}
\ee
for $|t|\gg 4m^2$.
Thus the vertex correction contains an infrared–sensitive imaginary term when
soft-photon emission is omitted.  In the standard Fock-space formulation this
imaginary contribution does not represent a physical instability but simply
signals that the truncated amplitude does not yet include real soft-photon
radiation.  This point was already emphasized in the early analysis of
Brown and Feynman \cite{Brown}, who showed that the apparent
imaginary part originates from computing the virtual amplitude without the
accompanying soft-photon emission.

Once soft emissions up to an energy cutoff $\omega_{\max}$ are incorporated, the
imaginary part cancels in the inclusive probability, as required by the
Bloch–Nordsieck/KLN mechanism.  In this sense the infrared structure reflects
the universal phase coherence of soft photons rather than any stochastic
dynamics of the electromagnetic field.

If we include explicitly the soft photons whose maximum energy is set to
$\omega_{\max}$, then Eq.~\bref{f} becomes (see Section~122 of
\cite{Berestetskii1})
\be
f_{\omega_{max}}=1-\frac{\alpha}{2}F\left(\frac{|q|}{2m}\right)
\ln\frac{m}{2\omega_{max}}+\frac{\alpha}{2}F_1+\alpha F_2.
\ee
Here $F$ is the Spence's function defined by
\be
F(\xi)=\int_0^\xi\frac{\ln (1+x)}{x}dx
\ee
and
\be
F(\xi)=\frac{2}{\pi}\left[\frac{2\xi^2+1}{\xi\sqrt{(\xi^2+1)}}
\ln(\xi+\sqrt{(\xi^2+1)})-1\right].
\ee
Also
\be
F_1=\frac{2\epsilon}{\pi|{\bf p}|}\ln\frac{\epsilon+|{\bf p}|}{m}
-\frac{2m^2+|{\bf q}|^2}{\pi\epsilon^2}\int_0^1\frac{dx}{a\sqrt{1-a}}
\ln\frac{1+\sqrt{1-a}}{\sqrt{a}}
\ee
with
\be
p^\mu=(\epsilon, {\bf p})~\mbox{and}~a=\frac{1}{\epsilon^2}
[m^2+{\bf q}^2(1-x)],
\ee
and $F_2$ is defined by
\be
f_\lambda(-{\bf q}^2)=1-\frac{1}{2}\alpha F(|{\bf q}|/(2m))\ln(m/\lambda)
+\alpha F_2.
\ee
In the limit of non-relativistic approximation, 
\be
f_{\omega_{max}}=1-\frac{\alpha q^2}{3\pi m^2}\left(
\ln\frac{m}{2\omega_{max}}+\frac{11}{24}\right)
\qquad (q^2\ll m^2).
\ee

Thus there appears neither imaginary part nor IRD in this standpoint.

\item {\bf Photon Propagator}

The last comment is concerned with the generalized fluctuation–dissipation
theorem of the electromagnetic fields (see Section 76 of
\cite{Berestetskii1} and \cite{Callen}),
\be
\left(A_i^{(1)}A_k^{(2)}\right)_\omega
=-\coth\frac{\hbar\omega}{2T}\,\mbox{Im} D_{ik}^R(\omega ;{\bf r}_1,{\bf r}_2).
\label{fluc1}
\ee
Here $(x_ax_b)_\omega$ indicates the Fourier component of
$\frac{1}{2}\langle x_a(t)x_b(0)+x_b(0)x_a(t)\rangle$ with respect
to time and $\mbox{Im}D_{ik}^R$ is the imaginary part of the retarded
photon propagator.
\begin{figure}[h]
\begin{center}
\includegraphics[scale=0.6]{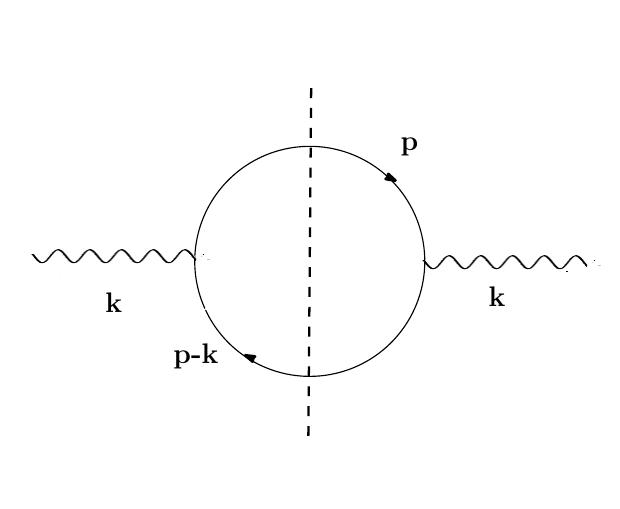}
\caption{Self energy diagram of photon.}
\label{fig:SEP}
\end{center}
\end{figure}
This comes from Kubo formula \cite{Kubo} on the generalized susceptibilities
$\alpha_{ik}$,
\be
\alpha_{ik}(\omega)=\frac{i}{\hbar}\int_0^\infty e^{i\omega t}
\langle A_i(t)A_k(0)-A_k(0)A_i(t)\rangle dt
\ee
and 
\be
\left(A_i^{(1)}A_k^{(2)}\right)_\omega=\frac{i\hbar}{2}
\coth\frac{\hbar\omega}{2T}\left[\alpha_{ki}^*(\omega;{\bf r}_2,{\bf r}_1)
-\alpha_{ik}(\omega;{\bf r}_1,{\bf r}_2)\right].
\ee
In the $T\to 0$ limit, the retarded Green function is reduced to the
Feynman Green function by
\bea
D_{ik}(\omega, {\bf k})&=&\mathrm{Re}\, D_{ik}^R(\omega, {\bf k})
+i\,\mathrm{sign}(\omega)\,\mathrm{Im}\, D_{ik}^R(\omega, {\bf k})
\nonumber\\
&=&\frac{4\pi\hbar}{\omega^2/c^2-k^2+i0}
\left(\delta_{ik}-\frac{c^2k_i k_k}{\omega^2}\right).
\eea
Thus we obtain 
\be
\left(A_i^{(1)}A_k^{(2)}\right)_\omega=\mbox{Im} D_{ik}(\omega ;{\bf r}_1,{\bf r}_2),
\label{fluc2}
\ee
in the $T\to 0$ limit.
The exact propagator $\mathcal{D}_{\mu\nu}$ is written in terms of the exact
polarization $\mathcal{P}_{\mu\nu}$ as
\be
\mathcal{D}_{\mu\nu}=\mathcal{D}_{\mu\nu}^{(0)}+\mathcal{D}_{\mu\lambda}^{(0)}
\frac{\mathcal{P}^{\lambda\rho}}{4\pi}\mathcal{D}_{\rho\nu}.
\ee
Using $\mathcal{P}_\mu^\mu=3\mathcal{P}$, we know that
\be
\mathcal{P}(t)=\frac{\alpha}{3\pi}t\left(\ln\frac{t}{m^2}-i\pi\right)
\qquad (t\gg 4m^2),
\ee
with $t=k^2$.
Thus the fluctuation of $A_i$ appears in the type of Fig.~3 and is not
concerned with IRD.

\end{itemize}
A crucial conceptual point should be emphasized at this stage.
The infrared fluctuations appearing in the Fock--space formulation of QED
do not represent stochastic classical noise.
Rather, they manifest themselves as coherent phase fluctuations encoded in a
unitary transformation acting on the charged-particle state.
The soft-photon cloud accompanying a charged particle modifies the state as
\begin{equation}
|\Psi\rangle \;\longrightarrow\; e^{i\Phi_{\rm IR}}\,|\Psi\rangle ,
\end{equation}
where the infrared phase $\Phi_{\rm IR}$ is universal and fixed by gauge
invariance.
As a result, the quantum state remains pure and fully coherent, despite the
presence of infrared-sensitive contributions.

This distinction is essential.
Genuine stochastic dynamics would require an irreversible loss of phase
information and a corresponding mixing of the density matrix.
In contrast, the infrared sector of QED is governed by coherent unitary
dressing, and its fluctuations cannot be interpreted as classical random
noise.

\section{Infrared cancellation and detector resolution}

The analysis in the previous section shows how infrared singularities
arise in the virtual sector through the logarithmic dependence on the
regulated photon mass $\lambda$.  However, the Fock-space formulation
also provides a complete mechanism for the cancellation of these
divergences once real soft-photon emission is included.

We may summarize the one-loop virtual correction as
\be
\sigma_{\rm virtual}
=
\sigma_0\left[
1-\frac{\alpha}{\pi} L 
\ln\!\left(\frac{\lambda}{m}\right)
+ \cdots
\right],
\label{virt-app-main}
\ee
where $L$ is the universal eikonal factor determined by the external
kinematics:
\begin{equation}
L=\ln\!\left(\frac{2p_+\cdot p_-}{m^{2}}\right)-1 .
\end{equation}

Soft real photon emission with energy $0<\omega<\omega_{\max}$ gives
\be
\frac{{\rm d}\sigma_{\rm soft}}{{\rm d}\omega}
=
\sigma_0\,\frac{\alpha}{\pi}\,L\,\frac{1}{\omega},
\qquad
0<\omega<\omega_{\max},
\label{soft-spectrum-main}
\ee
and integrates to
\be
\sigma_{\rm soft}(\omega_{\max})
=
\sigma_0\,
\frac{\alpha}{\pi}L
\ln\!\left(\frac{\omega_{\max}}{\lambda}\right)
+ \cdots .
\label{soft-app-main}
\ee

Adding Eqs.~\bref{virt-app-main} and \bref{soft-app-main} yields the
inclusive cross section
\be
\sigma_{\rm incl}(\omega_{\max})
=
\sigma_{\rm virtual}+\sigma_{\rm soft}(\omega_{\max})
=
\sigma_0\left[
1+\frac{\alpha}{\pi}L
\ln\!\left(\frac{\omega_{\max}}{m}\right)
+ \cdots
\right],
\label{incl-main}
\ee
in which the regulator $\lambda$ cancels exactly.  This is the standard
Bloch--Nordsieck/KLN mechanism: the same factor $L$ appears in both
virtual and real soft-photon contributions, enforced by gauge invariance
and the Ward--Takahashi identity.
It is important to distinguish detector coarse-graining from genuine
stochastic dynamics.
A finite detector resolution implies that sufficiently soft photons are not
individually resolved, and one must sum inclusively over unobserved modes.
However, this coarse-graining does not destroy the phase correlations between
virtual and real infrared contributions that are required for the
Bloch--Nordsieck and Kinoshita--Lee--Nauenberg cancellations.

In particular, the infrared finiteness of inclusive observables relies on
the persistence of coherent phase information across amplitudes.
This situation differs fundamentally from classical stochastic dynamics,
where coarse-graining leads to an irreversible averaging over random phases.
Detector resolution therefore does not induce classicalization of the
infrared sector, but merely reflects the operational definition of physical
observables.
Schematically, stochastic dynamics corresponds to an ensemble average
\begin{equation}
\rho \;\longrightarrow\; \int d\xi\, P(\xi)\,
U(\xi)\,\rho\,U^\dagger(\xi),
\end{equation}
whereas infrared effects in QED are described by a single universal unitary
transformation,
\begin{equation}
\rho \;\longrightarrow\; U_{\rm IR}\,\rho\,U_{\rm IR}^\dagger .
\end{equation}

The parameter $\omega_{\max}$ is not a physical cutoff; i
The parameter $\omega_{\max}$ is not a physical cutoff; it is only a
theoretical separator delimiting the region in which the eikonal
approximation holds.  Changing $\omega_{\max}$ simply reshuffles the
probability between ``soft'' and ``hard'' real emission and has no direct
observable meaning.

The experimentally relevant quantity is the cross section defined with the
detector energy resolution $\Delta E$.  If the electron loses an energy
less than $\Delta E$, the event is experimentally indistinguishable from
the Born process.  Hence the physical cross section is
\be
\sigma_{\rm phys}(\Delta E)
=
\sigma_0\left[
1+\frac{\alpha}{\pi}L
\ln\!\left(\frac{\Delta E}{m}\right)
+ \cdots
\right].
\label{sigmaphys-main}
\ee
If $\Delta E$ is small, soft-photon emission is experimentally resolved
and $\sigma_{\rm phys}\to \sigma_0$.  If $\Delta E$ is large, more
soft-photon radiation remains unresolved and the deviation from $\sigma_0$
becomes finite and physically meaningful.  The formal limit $\Delta E\to 0$
would correspond to resolving all soft photons individually, which is
experimentally impossible; therefore no divergence occurs in any
measurable cross section.  This clarifies the physical meaning of the
imaginary parts encountered in Sec.~2: they reflect the neglect of real
soft radiation and are removed by the inclusive treatment.

\section{Criterion for stochastic interpretation and its failure in
four-dimensional Maxwell theory}
\label{sec:stochastic}

The appearance of an imaginary part in the Schwinger--Keldysh (SK) effective
action is often taken as an indication of stochastic dynamics.
However, this feature alone is not sufficient to justify a physical
interpretation in terms of classical noise.
A genuine stochastic description requires additional dynamical conditions:
long-wavelength modes must undergo infrared growth and effective
classicalization, leading to an irreversible loss of phase information.
Only under these conditions can the imaginary part of the SK effective action
be consistently interpreted as a noise kernel, and a Hubbard--Stratonovich (HS)
auxiliary field be identified with a stochastic force.

These conditions are realized for nearly massless scalar fields in
de~Sitter spacetime.
In that case, long-wavelength modes continuously accumulate outside the Hubble
radius, and the imaginary part of the SK effective action exhibits secular
infrared growth.
The HS-transformed auxiliary field then acquires a natural interpretation as a
classical stochastic force, reproducing the well-known classicalization of
super-Hubble scalar fluctuations.
This mechanism underlies the stochastic inflation picture for light scalar
fields.

Four-dimensional Maxwell theory, however, violates the above criterion in a
fundamental way.
Because the Maxwell action is conformally invariant in four dimensions, the
infrared spectrum of electromagnetic fields does not grow in a de~Sitter
background.
Long-wavelength photon modes do not freeze, no accumulation of infrared power
occurs, and the electromagnetic field remains intrinsically quantum coherent.
As a result, although the SK effective action of QED formally admits the same
HS decomposition, the auxiliary field $\xi$ does not acquire the status of a
physical stochastic force.

This conclusion is fully consistent with the Fock--space analysis presented in
Secs.~2 and~3.
In QED the infrared sector is governed by gauge invariance and by the coherent
phase structure of soft photons, as enforced by the Ward--Takahashi identity.
The cancellation of infrared divergences relies on correlated phases between
virtual and real emissions, and detector coarse-graining does not induce
irreversible averaging over random fluctuations.
The formal similarity between the SK/HS representations of QED and of scalar
field theory in de~Sitter spacetime therefore does not imply a common physical
infrared behavior.

We thus conclude that, for four-dimensional Maxwell theory, the conditions
required for a stochastic interpretation of the SK effective action are never
met.
Infrared effects in QED are governed by coherent quantum dressing rather than
by classical stochastic dynamics.



\section{Conclusion}

In this work we have established a no-go result for a stochastic
interpretation of infrared effects in quantum electrodynamics.
Using the standard Fock--space formulation, we have shown that infrared
fluctuations in QED manifest themselves as coherent phase correlations,
enforced by gauge invariance, rather than as classical stochastic noise.
The universal soft-photon dressing accompanying charged particles preserves
quantum coherence and guarantees the
Bloch--Nordsieck/Kinoshita--Lee--Nauenberg cancellation for all inclusive
observables.
Infrared divergences of perturbative amplitudes therefore do not signal a
physical instability of the theory, but reflect the necessity of accounting
for this coherent dressing.

We have further clarified the role of detector resolution in infrared
physics.
The finite energy resolution $\omega_{\max}$ defines which soft photons are
experimentally unresolved and must be summed over inclusively, but it does
not induce classicalization of the infrared sector.
Detector coarse-graining integrates over unobserved modes while preserving
the phase correlations required for infrared cancellation.
This situation is fundamentally different from genuine stochastic dynamics,
where irreversible averaging over random phases leads to classical noise.

Within the Schwinger--Keldysh framework, the appearance of an imaginary part
in the effective action is often taken as an indication of stochastic
behavior.
We have shown that this feature alone is insufficient.
A physical Langevin interpretation requires infrared decoherence driven by
dynamical classicalization of long-wavelength modes.
Such conditions are realized for nearly massless scalar fields in de~Sitter
spacetime, but are never satisfied for four-dimensional Maxwell theory.
Because of conformal invariance, electromagnetic modes do not exhibit
infrared growth or secular accumulation, and soft photons remain quantum
coherent.

As a consequence, although the SK effective action of QED formally admits a
Hubbard--Stratonovich representation, the associated auxiliary field cannot be
interpreted as a physical stochastic force.
The formal similarity between the SK/HS structures of QED and of scalar field
theory in de~Sitter spacetime therefore does not imply a common physical
infrared behavior.
For Maxwell theory, infrared effects are governed by coherent quantum
dressing protected by gauge invariance, not by infrared-induced classical
stochastic dynamics.

\bibliographystyle{unsrt}

\end{document}